\documentclass[12pt]{article}
\usepackage{epsfig}
\usepackage{amssymb}
\usepackage{amsmath}
\usepackage{amsfonts}

\newcommand{\be}{\begin{equation}}
\newcommand{\ee}{\end{equation}}
\newcommand{\bea}{\begin{eqnarray}}
\newcommand{\eea}{\end{eqnarray}}

\newcommand{\kt}{\rangle}
\newcommand{\br}{\langle}

\newcommand{\ed}{\end{document}}
\newcommand{\bbr}{\br\!\br}

\newcommand{\bi}{\begin{itemize}}
\newcommand{\ei}{\end{itemize}}

\begin{document}

\title{Comment on ``Time-dependent quasi-Hermitian Hamiltonians
and the unitary\\ quantum evolution''}
\author{\\
Ali Mostafazadeh
\\
\\
Department of Mathematics, Ko\c{c} University,\\
34450 Sariyer, Istanbul, Turkey\\ amostafazadeh@ku.edu.tr}
\date{ }
\maketitle

\begin{abstract}

In arXiv:0710.5653v1 M.~Znojil claims that he has found and
corrected an error in my paper: [Phys.~Lett.~B \textbf{650}, 208
(2007), arXiv:0706.1872v2] and that it is possible to escape its
main conclusion, namely that the unitarity of the time-evolution
and observability of the Hamiltonian imply time-independence of
the metric operator. In this note I give a very short calculation
showing that the analysis given by M.~Znojil also leads to the
same conclusion as of my above-mentioned paper. This is actually a
reconfirmation of the validity of the results of the latter paper.

\vspace{5mm}

\noindent PACS number: 03.65.Ca, 11.30.Er, 03.65.Pm,
11.80.Cr\vspace{2mm}


\end{abstract}


First I recall the notation used in \cite{znojil}.
    \begin{itemize}
    \item $\Theta$ is a possibly time-dependent (positive) metric
    operator.
    \item $H$ is a possibly time-dependent $\Theta$-pseudo-Hermitian
    Hamiltonian operator acting on a reference Hilbert space
    ${\cal H}$ with the inner product $\br\cdot|\cdot\kt$, i.e.,
        \be
        H(t)^\dagger=\Theta(t)H(t)\Theta(t)^{-1}.
        \label{ph}
        \ee
    \item $\omega:=\sqrt\Theta$ and $h:=\omega H\omega^{-1}$ is the
    equivalent Hermitian Hamiltonian with evolution operator $u$, i.e.,
    $i\hbar\partial_t u(t)=h(t)u(t)$ and $u(0)=I$,
    where $I$ stands for the identity operator.
    \item $U_R$ is the evolution operator for $H$, i.e.,
        \be
        i\hbar\partial_t U_R(t)=H(t)U_R(t),~~~~U_R(0)=I.
        \label{U=}
        \ee
    \end{itemize}
According to the first of Eqs.~(17) of \cite{znojil},
        \be
        U_R(t)=\omega(t)^{-1}u(t)\omega(0).
        \label{e17}
        \ee
Because $h(t)$ is Hermitian, $u(t)$ is unitary, i.e.,
$u(t)^\dagger=u(t)^{-1}$. Using this observation, the fact that
$\omega(t)$ is Hermitian, and Eqs.\ $\omega(t)^2=\Theta(t)$ and
(\ref{e17}), we have
    \[U_R(t)^{-1\dagger}\Theta(0)\:U_R^{-1}=
    \omega(t)u(t)\omega(0)^{-1}\Theta(0)\:\omega(0)^{-1}u(t)^{-1}
    \omega(t)=\Theta(t).\]
Given that in Ref.~\cite{plb} I use $\xi_+$ for $\Theta$, this
equation, is identical to Eq.~(11) of \cite{plb}. As explained in
\cite{plb}, if we use this equation to compute
$\partial_t\Theta(t)$ and employ (\ref{U=}), we find Eq.~(12) of
\cite{plb}, namely
    \[H(t)^\dagger=\Theta(t)H(t)\Theta(t)^{-1}
    -i\Theta(t)\frac{d}{dt}\Theta(t)^{-1}.\]
This together with (\ref{ph}) and the invertibility of $\Theta(t)$
imply
    \be
    \frac{d}{dt}\Theta(t)=0.
    \label{final}
    \ee
Therefore, contrary to the claim made by M.~Znojil in
\cite{znojil}, the metric operator is indeed constant, and there
is no error in \cite{plb}. The root of the misjudgment made in
\cite{znojil} seems to be the rather deceptive nature of the
notation $\bbr\cdot|$ that hides the restriction imposed by
unitarity on the metric operator in Eq.~(18) of \cite{znojil}.

\ed